\documentclass[iop]{emulateapj}
\pdfoutput=1

\newcommand{\myemail}{josemari@alumni.nd.edu}

\newcommand{\host}              {1}
\newcommand{\z}                 {2}
\newcommand{\IRAC}              {3}
\newcommand{\K}                 {4}
\newcommand{\mass}              {5}
\newcommand{\UV}                {6}
\newcommand{\SFR}               {7}

\defcitealias{bloom01}          {1}
\defcitealias{bloom98}          {2}
\defcitealias{djorgovski01}     {3}
\defcitealias{bloom99}          {4}
\defcitealias{tinney98}         {5}
\defcitealias{djorgovski03}     {6}
\defcitealias{djorgovski98}     {7}
\defcitealias{christensen05}    {8}
\defcitealias{bloom03}          {9}
\defcitealias{le floc'h02}     {10}
\defcitealias{piro02}          {11}
\defcitealias{price02a}        {12}
\defcitealias{price02b}        {13}
\defcitealias{price03}         {14}
\defcitealias{barth03}         {15}
\defcitealias{jakobsson05b}    {16}
\defcitealias{vreeswijk06}     {17}
\defcitealias{maiorano06}      {18}
\defcitealias{hjorth03}        {19}
\defcitealias{jakobsson04}     {20}
\defcitealias{rau05}           {21}
\defcitealias{prochaska04}     {22}
\defcitealias{wiersema04}      {23}
\defcitealias{soderberg06}     {24}
\defcitealias{pellizza06}      {25}
\defcitealias{foley05}         {26}
\defcitealias{pian06}          {27}
\defcitealias{colless01}       {28}
\defcitealias{della valle06a}  {29}
\defcitealias{chary02}         {30}
\defcitealias{castro cerón06}  {31}
\defcitealias{le floc'h06}     {32}
\defcitealias{vreeswijk99}     {33}
\defcitealias{le floc'h03}     {34}
\defcitealias{gorosabel03a}    {35}
\defcitealias{gorosabel03b}    {36}
\defcitealias{masetti05}       {37}
\defcitealias{wainwright07}    {38}
\defcitealias{gorosabel05a}    {39}
\defcitealias{gorosabel05b}    {40}
\defcitealias{rau04}           {41}
\defcitealias{malesani04}      {42}
\defcitealias{della valle06b}  {43}
\defcitealias{kocevski07}      {44}
\defcitealias{thöne08}         {45}
\def \privatehjorth            {46}
\defcitealias{cobb06}          {47}
\defcitealias{christensen04}   {48}
\defcitealias{bloom02}         {49}
\defcitealias{michałowski09}   {50}
\defcitealias{margutti07}      {51}
\defcitealias{sollerman06}     {52}
\defcitealias{mangano07}       {53}

\shorttitle{$M_\star$ distribution in GRB hosts}
\shortauthors{Castro Cer\'on et al.}

\begin{document}

\title{On the distribution of stellar masses in gamma-ray burst host galaxies\altaffilmark{1}}

\author{J. M. Castro Cer\'on\altaffilmark{2,3},
        M. J. Micha{\l}owski\altaffilmark{2,4},
           J. Hjorth\altaffilmark{2},
           D. Malesani\altaffilmark{2},
           J. Gorosabel\altaffilmark{5}, \\
           D. Watson\altaffilmark{2},
     J. P. U. Fynbo\altaffilmark{2},
       and M. Morales Calder\'on\altaffilmark{6}
       }

\altaffiltext{1}{This work is based in part on observations made with the {\em Spitzer Space Telescope}, which is operated by the Jet Propulsion Laboratory, California Institute of Technology, under a contract with NASA.}
\altaffiltext{2}{Dark Cosmology Centre, Niels Bohr Institute, University of Copenhagen, Juliane Maries Vej 30, DK-2100 Copenhagen \O, Denmark.}
\altaffiltext{3}{Herschel Science Centre (ESAC/ESA), Camino Bajo del Castillo, s/n, E-28.692 Villanueva de la Ca\~nada (Madrid), Spain; \myemail.}
\altaffiltext{4}{Scottish Universities Physics Alliance, Institute for Astronomy, University of Edinburgh, Royal Observatory, Edinburgh, EH9 3HJ, UK.}
\altaffiltext{5}{Instituto de Astrof\'{\i}sica de Andaluc\'{\i}a (CSIC), Glorieta de la Astronom\'{\i}a, s/n, E-18.008 Granada, Spain.}
\altaffiltext{6}{{\em Spitzer} Science Center, California Institute of Technology, Pasadena CA 91125, USA.}

\begin{abstract}

We analyse {\em Spitzer} images of 30 long-duration gamma-ray burst (GRB) host galaxies. We estimate their total stellar masses ($M_\star$) based on the rest-frame $K$-band luminosities ($L_{K_{\mathrm{rest}}}$) and constrain their star formation rates (SFRs, not corrected for dust extinction) based on the rest-frame UV continua. Further, we compute a mean $M_\star/L_{K_{\mathrm{rest}}}$ = 0.45\,$M_\sun$/$L_\sun$. We find that the hosts are low $M_\star$, star-forming systems. The median $M_\star$ in our sample ($\langle M_\star \rangle$ = 10$^{9.7}$\,$M_\sun$) is lower than that of ``field'' galaxies (e.g., Gemini Deep Deep Survey). The range spanned by $M_\star$ is 10$^7$\,$M_\sun$ $<$ $M_\star$ $<$ 10$^{11}$\,$M_\sun$, while the range spanned by the dust-uncorrected UV SFR is 10$^{-2}$\,$M_\sun$\,yr$^{-1}$ $<$ SFR $<$ 10\,$M_\sun$\,yr$^{-1}$. There is no evidence for intrinsic evolution in the distribution of $M_\star$ with redshift. We show that extinction by dust must be present in at least 25\% of the GRB hosts in our sample and suggest that this is a way to reconcile our finding of a relatively lower UV-based, specific SFR ($\phi$ $\equiv$ SFR/$M_\star$) with previous claims that GRBs have some of the highest $\phi$ values. We also examine the effect that the inability to resolve the star-forming regions in the hosts has on $\phi$.

\end{abstract}


\section{Introduction}
  \label{intro}

It is central to contemporary cosmology to map the buildup of cosmic structure and star formation (SF), and we know that the detection of a gamma-ray burst (GRB) is an indication that its host galaxy harbours massive SF. GRBs are pulses of $\gamma$-rays from sources of cosmological origins and are the most luminous, photon-emitting events in the universe. As tracers of SF, they have some fundamental advantages: dust extinction has essentially no effect in their detection at $\gamma$-ray and X-ray wavelengths and GRBs can be observed to very high redshifts. That is to say that GRBs can furnish us with unique eyes to gainfully look at the star-forming universe. But the following critical questions should be answered in order to use GRBs as tracers of SF: What is the level of bias in those GRB host samples that have been optically selected? And what is the intrinsic bias in the GRB-SF rate?

A canonical model is rather well established for long-duration GRBs; they occur in star-forming regions within star-forming galaxies \citep{bloom02,gorosabel03a,christensen04,fruchter06} and are associated with stellar core collapse events and hence with high-mass SF \citep[e.g.][]{galama98,hjorth03,stanek03,zeh04,campana06}. The emerging picture, however, is a complex one. Most GRB host galaxies are faint and blue \citep{fruchter99,le floc'h03}. A few hosts show tentative evidence of very high star formation rates \citep[SFRs;][]{chary02,berger03,michałowski08}, but their optical properties do not appear typical of the galaxies that can be found in blind submillimeter galaxy surveys \citep{tanvir04,fruchter06}.

It is currently debated how GRB hosts relate to other known populations of star-forming galaxies. At redshifts around 3, the UV luminosities of host galaxies and the metallicities of GRB sightlines are consistent with the expectations if hosts are drawn from the underlying population of all star-forming galaxies weighted with the total SF density per luminosity bin \citep{jakobsson05a,fynbo08}. With {\em Spitzer}'s \citep{werner04} IRAC \citep[Infrared Array Camera;][]{fazio04} mid-infrared (MIR) photometry, together with optical and near-infrared (NIR) data, we can establish how the host galaxies relate to other star-forming populations in terms of total stellar mass ($M_\star$). This is essential if we are to understand the full range of properties of star-forming galaxies at high redshifts and fully exploit the potential of GRBs as probes of cosmic SF.

\citet{castro cerón06} studied a sample of six long-duration GRB host galaxies observed with IRAC and MIPS \citep[Multiband Imager Photometer for {\em Spitzer}\/;][]{rieke04}. They estimated their $M_\star$ based on rest-frame $K$-band luminosity densities and constrained their SFRs based on the entire available spectral energy distribution (SED). In this work, we extend the computations to a sample of 30 but constrain only the dust-uncorrected UV SFRs with the rest-frame UV continuum. This larger sample ought to allow for a more robust statistical analysis, as well as to probe the distribution of $M_\star$ in redshift space. To determine $M_\star$ we utilise rest-frame $K$ flux densities (interpolated from observed IRAC and NIR fluxes). This extends the data set presented by \citet{castro cerón06}, yielding accurate values of $M_\star$ in a large host galaxy sample. To determine the dust-uncorrected UV SFRs, we use rest-frame UV flux densities (interpolated from observed optical fluxes). These dust-uncorrected UV SFRs are lower limits to the total SFR of a galaxy due to the possible extinction by dust, and we compare them with those of \citet{castro cerón06}. Our paper is organised as follows. An overview of the sample selection is given in Section \ref{data}. The analytic methodology is described in Section \ref{method}. $M_\star$ for the sample are derived in Section \ref{M*}, and Section \ref{SFRs} sees the computation of the dust-uncorrected UV SFRs. We conclude in Section \ref{analysis} with analysis and discussion. We assume an $\Omega_m$ = 0.3, $\Omega_\Lambda$ = 0.7 cosmology with $H_0$ = 70\,km s$^{-1}$ Mpc$^{-1}$.

\begin{deluxetable*}{lcccccccccccc}
    \setlength{\tabcolsep}{0.03in}
\tabletypesize{\footnotesize}
   \tablewidth{0pt}
 \tablecaption{Hosts: Flux Densities, Total Stellar Masses, and Star Formation Rates
        \label{flux:m*:sfr}
              }

\tablehead{
                                                  \colhead{}                             & \multicolumn{2}{c}{Redshift}                               & \colhead{} & \multicolumn{2}{c}{IRAC}                              & \colhead{} & \multicolumn{2}{c}{$K_{\rm rest}$ (21\,980\,\AA)}                                                             & \colhead{}                   & \multicolumn{2}{c}{UV$_{\rm rest}$ (2\,800\,\AA)}                                                & \colhead{}                      \\
\cline{2-3} \cline{5-6} \cline{8-9} \cline{11-12} \colhead{}                             & \colhead{}                  & \colhead{}                   & \colhead{} & \colhead{}                            & \colhead{}    & \colhead{} & \colhead{}                          & \colhead{}                                                              & \colhead{$M_{\star (K)}$}    & \colhead{}                             & \colhead{}                                              & \colhead{SFR$_{\rm (UV)}$}      \\
                                                  \colhead{GRB Host}                     & \colhead{$z$}               & \colhead{Refs.}              & \colhead{} & \colhead{$f_\nu$ ($\mu$Jy)}           & \colhead{Ch.} & \colhead{} & \colhead{$f_\nu$ ($\mu$Jy)}         & \colhead{Refs.}                                                         & \colhead{(10$^9$\,$M_\sun$)} & \colhead{$f_\nu$ ($\mu$Jy)}            & \colhead{Refs.}                                         & \colhead{($M_\sun$\,yr$^{-1}$)} \\
                                                  \colhead{(\host)}                      & \multicolumn{2}{c}{(\z)}                                   & \colhead{} & \multicolumn{2}{c}{(\IRAC)}                           & \colhead{} & \multicolumn{2}{c}{(\K)}                                                                                      & \colhead{(\mass)}            & \multicolumn{2}{c}{(\UV)}                                                                        & \colhead{(\SFR)}
          }

\startdata
                                    \object[GRB970228] {970228}                   ...... &       0.70                  & \citetalias{bloom01}         &            &   $<$3.7                              &  1            &            &   $<$4.2                            & $\ddagger$,                  \citetalias{chary02}                       &   $<$5.7                     &       0.34  $\pm$   0.16               & \citetalias{christensen04}                              &       0.60  $\pm$ 0.28        \\
                                    \object[GRB970508] {970508}                   ...... &       0.83                  & \citetalias{bloom98}         &            &   $<$2.1\tablenotemark{a}             &  2            &            &   $<$1.8                            & \citetalias{castro cerón06}, \citetalias{chary02}                       &   $<$3.5                     &       0.28  $\pm$   0.15               & \citetalias{christensen04}                              &       0.71  $\pm$ 0.38        \\
                                    \object[GRB970828] {970828}                   ...... &       0.96                  & \citetalias{djorgovski01}    &            &      3.9 $\pm$   0.3\tablenotemark{a} &  2            &            &      3.7 $\pm$   0.3                & \citetalias{castro cerón06}, \citetalias{djorgovski01}                  &      9.5   $\pm$ 0.9         &    $<$0.44                             & \citetalias{le floc'h03}                                &    $<$1.5                     \\
                                    \object[GRB980326] {980326}                   ...... & $\sim$1.0                   & \citetalias{bloom99}         &            &   $<$2.7                              &  2            &            &   $<$2.6                            & $\ddagger$,                  \citetalias{chary02}                       &   $<$7.1                     &    $<$0.015                            & \citetalias{bloom02}                                    &    $<$0.056                   \\
                                    \object[GRB980425] {980425}                   ...... &       0.0085                & \citetalias{tinney98}        &            & 2\,977   $\pm$ 101                    &  2            &            & 6\,389   $\pm$ 395                  & $\ddagger$\tablenotemark{b}, \citetalias{le floc'h06}                   &      1.1   $\pm$ 0.1         &  1\,748     $\pm$ 173                  & $\ddagger$\tablenotemark{c}, \citetalias{michałowski09} &       0.39  $\pm$ 0.04        \\
                                    \object[GRB980613] {980613}                   ...... &       1.10                  & \citetalias{djorgovski03}    &            &     38   $\pm$   1\tablenotemark{a}   &  2            &            &     42   $\pm$   1                  & \citetalias{castro cerón06}, \citetalias{djorgovski03}                  &    142     $\pm$ 3           &       0.83  $\pm$   0.11               & \citetalias{djorgovski03},   \citetalias{chary02}       &       3.6   $\pm$ 0.5         \\
                                    \object[GRB980703] {980703}                   ...... &       0.97                  & \citetalias{djorgovski98}    &            &     11   $\pm$   1\tablenotemark{a}   &  2            &            &     11   $\pm$   1                  & \citetalias{castro cerón06}, \citetalias{vreeswijk99}                   &     29     $\pm$ 2           &       3.2   $\pm$   0.1                & \citetalias{christensen04}                              &      10.9   $\pm$ 0.3         \\
                                    \object[GRB981226] {981226}                   ...... &       1.11                  & \citetalias{christensen05}   &            &      4.5 $\pm$   0.5\tablenotemark{a} &  2            &            &      4.6 $\pm$   0.5                & \citetalias{castro cerón06}, \citetalias{christensen05}                 &     16     $\pm$ 2           &       0.27  $\pm$   0.03               & \citetalias{christensen05}                              &       1.2   $\pm$ 0.1         \\
                                    \object[GRB990506] {990506}                   ...... &       1.31                  & \citetalias{bloom03}         &            &      2.0 $\pm$   0.7                  &  2            &            &      2.0 $\pm$   0.8                & $\ddagger$,                  \citetalias{le floc'h03}                   &      9.3   $\pm$ 3.8         &       0.20  $\pm$   0.04               & \citetalias{le floc'h03}                                &       1.2   $\pm$ 0.2         \\
                                    \object[GRB990705] {990705}                   ...... &       0.84                  & \citetalias{le floc'h02}     &            &     19   $\pm$   1\tablenotemark{a}   &  2            &            &     18   $\pm$   1                  & \citetalias{castro cerón06}, \citetalias{le floc'h02}                   &     36     $\pm$ 2           & $\sim$1.8   $\pm$   0.3                & \citetalias{le floc'h02}                                & $\sim$4.7   $\pm$ 0.8         \\
                                    \object[GRB000210] {000210}                   ...... &       0.85                  & \citetalias{piro02}          &            &      3.3 $\pm$   2.0                  &  2            &            &      3.2 $\pm$   1.8                & $\ddagger$,                  \citetalias{gorosabel03a}                  &      6.4   $\pm$ 3.6         &       0.79  $\pm$   0.07               & \citetalias{christensen04}                              &       2.1   $\pm$ 0.2         \\
                                    \object[GRB000418] {000418}                   ...... &       1.12                  & \citetalias{bloom03}         &            &      4.8 $\pm$   1.8                  &  2            &            &      5.0 $\pm$   1.9                & $\ddagger$,                  \citetalias{gorosabel03b}                  &     17     $\pm$ 7           &       1.33  $\pm$   0.04               & \citetalias{christensen04}                              &       6.1   $\pm$ 0.2         \\
                                    \object[GRB000911] {000911}                   ...... &       1.06                  & \citetalias{price02a}        &            &   $<$4.3                              &  2            &            &   $<$4.3                            & $\ddagger$,                  \citetalias{masetti05}                     &  $<$13                       &       0.33  $\pm$   0.08               & \citetalias{masetti05}                                  &       1.4   $\pm$ 0.3         \\
                                    \object[GRB010921] {010921}                   ...... &       0.45                  & \citetalias{price02b}        &            &     11   $\pm$   2                    &  1            &            &     12   $\pm$   2                  & $\ddagger$,                  \citetalias{price02b}                      &      6.5   $\pm$ 0.9         &       2.2   $\pm$   0.1                & \citetalias{christensen04}                              &       1.6   $\pm$ 0.1         \\
                                    \object[GRB020405] {020405}                   ...... &       0.69                  & \citetalias{price03}         &            &   $<$5.4                              &  1            &            &   $<$5.3                            & $\ddagger$,                  \citetalias{wainwright07}                  &   $<$7.0                     &       2.1   $\pm$   0.1                & \citetalias{wainwright07}                               &       3.7   $\pm$ 0.2         \\
                                    \object[GRB020813] {020813}                   ...... &       1.26                  & \citetalias{barth03}         &            &   $<$2.5                              &  2            &            &   $<$2.6                            & $\ddagger$,                  \citetalias{wainwright07}                  &  $<$11                       &       0.41  $\pm$   0.08               & \citetalias{le floc'h03},    \citetalias{wainwright07}  &       2.3   $\pm$ 0.5         \\
                                    \object[GRB020819B]{020819B}                  ...... &       0.41                  & \citetalias{jakobsson05b}    &            &     97   $\pm$   2                    &  1            &            &    104   $\pm$   7                  & $\ddagger$,                  \citetalias{jakobsson05b}                  &     47     $\pm$ 3           &       4.3   $\pm$   2.6                & \citetalias{jakobsson05b}                               &       2.6   $\pm$ 1.5         \\
                                    \object[GRB021211] {021211}                   ...... &       1.01                  & \citetalias{vreeswijk06}     &            &   $<$2.2                              &  2            &            &   $<$2.2                            & $\ddagger$,                  \citetalias{wainwright07}                  &   $<$6.1                     &       0.20  $\pm$   0.04               & \citetalias{wainwright07}                               &       0.72  $\pm$ 0.15        \\
                                    \object[GRB030328] {030328}                   ...... &       1.52                  & \citetalias{maiorano06}      &            &  $<$29                                &  3            &            &  $<$27                              & $\ddagger$,                  \citetalias{gorosabel05a}                  & $<$170                       &       0.56  $\pm$   0.08               & \citetalias{gorosabel05a}                               &       4.6   $\pm$ 0.6         \\
                                    \object[GRB030329] {030329}                   ...... &       0.17                  & \citetalias{hjorth03}        &            &  $<$4.9                               &  1            &            &  $<$5.1                             & $\ddagger$,                  \citetalias{gorosabel05b}\tablenotemark{d} &   $<$0.37                    &       1.5   $\pm$   0.2                & \citetalias{gorosabel05b}                               &       0.14  $\pm$ 0.02        \\
                                    \object[GRB030429] {030429}\tablenotemark{e}  ...... &       2.66                  & \citetalias{jakobsson04}     &            &   $<$7.0                              &  4            &            &   $<$7.3                            & $\ddagger$,                  \citetalias{jakobsson04}                   & $<$124                       &    $<$0.060                            & \citetalias{jakobsson04}                                &    $<$1.3                     \\
                                    \object[GRB030528] {030528}\tablenotemark{e}  ...... &       0.78                  & \citetalias{rau05}           &            &   $<$4.6                              &  1            &            &   $<$3.8                            & $\ddagger$,                  \citetalias{rau04}                         &   $<$6.5                     &       7.2   $\pm$   1.4                & \citetalias{rau04}                                      &      16     $\pm$ 3           \\
                                    \object[GRB031203] {031203}                   ...... &       0.11                  & \citetalias{prochaska04}     &            &    216   $\pm$   3\tablenotemark{f}   &  1            &            &    192   $\pm$  13\tablenotemark{f} & $\ddagger$,                  \citetalias{malesani04}                    &      5.3   $\pm$ 0.4         &     119     $\pm$  39\tablenotemark{f} & \citetalias{margutti07}                                 &       4.3   $\pm$ 1.4         \\
                                    \object[GRB040924] {040924}                   ...... &       0.86                  & \citetalias{wiersema04}      &            &   $<$2.9                              &  1            &            &   $<$3.2                            & $\ddagger$,                  \citetalias{wainwright07}                  &   $<$6.5                     &    $<$1.1                              & \citetalias{wainwright07}                               &    $<$2.9                     \\
                                    \object[GRB041006] {041006}                   ...... &       0.72                  & \citetalias{soderberg06}     &            &   $<$2.9                              &  1            &            &   $<$3.1                            & $\ddagger$,                  \citetalias{wainwright07}                  &   $<$4.4                     &    $<$0.98                             & \citetalias{wainwright07}                               &    $<$1.8                     \\
                                    \object[GRB050223] {050223}                   ...... &       0.58                  & \citetalias{pellizza06}      &            &     18   $\pm$   2                    &  1            &            &     18   $\pm$   2                  & $\ddagger$,                  \citetalias{pellizza06}                    &     17     $\pm$ 1           &    $<$8.1                              & \citetalias{pellizza06}                                 &   $<$10                       \\
                                    \object[GRB050525A]{050525A}                  ...... &       0.61                  & \citetalias{foley05}         &            &   $<$1.6                              &  1            &            &   $<$1.6                            & $\ddagger$,                  \citetalias{della valle06b}                &   $<$1.6                     &    $<$0.48                             & \citetalias{della valle06b}                             &    $<$0.64                    \\
                                    \object[GRB060218] {060218}\tablenotemark{e}  ...... &       0.03                  & \citetalias{pian06}          &            & \nodata                               & \nodata       &            &     20   $\pm$   6                  & \citetalias{kocevski07}                                                 &      0.052 $\pm$ 0.015       &      15     $\pm$   3                  & \citetalias{sollerman06}                                &       0.053 $\pm$ 0.010       \\
                                    \object[GRB060505] {060505}                   ...... &       0.09\tablenotemark{g} & \citetalias{colless01}       &            & \nodata                               & \nodata       &            &    298   $\pm$  10                  & \citetalias{thöne08}                                                    &      5.8   $\pm$ 0.2         &      75     $\pm$   6                  & $\ddagger$\tablenotemark{c}, \citetalias{thöne08}       &       1.9   $\pm$ 0.2         \\
                                    \object[GRB060614] {060614}                   ...... &       0.13                  & \citetalias{della valle06a}  &            & \nodata                               & \nodata       &            &      3.8 $\pm$   0.7                &     \privatehjorth,          \citetalias{cobb06}                        &      0.15  $\pm$ 0.03        &       0.37  $\pm$   0.13               & \citetalias{mangano07}                                  &       0.019 $\pm$ 0.006       \\
  \enddata

\tablenotetext{a}   {Flux density values are taken from \citet{castro cerón06}; we refine their error estimates.}
\tablenotetext{b}   {Our photometry of 2MASS XSC Final Release (Two Micron All Sky Survey Extended Source Catalog; released 25 March 2003; \citealt{jarrett00}; {\tt http://www.ipac.caltech.edu/2mass/}), NIR ($K_{\rm S}$ band) archival data for galaxy \object[ESO 184- G 082]{ESO 184-G082} ($f_{21\,739\,\textrm{\tiny \AA}}$ = 6\,510\,$\mu$Jy $\pm$ 406\,$\mu$Jy).}
\tablenotetext{c}   {Our photometry of $GALEX$ ({\em Galaxy Evolution Explorer}; \citealt{martin03,martin05}; {\tt http://galex.stsci.edu/}), UV archival data for the host galaxies of GRBs \object[GRB980425]{980425} ($f_{2\,267\,\textrm{\tiny \AA}}$ = 1\,592\,$\mu$Jy $\pm$ 162\,$\mu$Jy) and \object[GRB060505]{060505} ($f_{2\,267\,\textrm{\tiny \AA}}$ = 72\,$\mu$Jy $\pm$ 10\,$\mu$Jy).}
\tablenotetext{d}   {$K$ band is the closest passband, blueward of IRAC, for which this host has data available in the literature. It is a poorly constrained upper limit. We make use of it nevertheless, for methodological consistency (see Section 3). But we note that in this particular case, given the low redshift of the host, a much closer representation of reality is provided by the lower limit $M_\star$ = 6.4 $\times$ 10$^7$\,$M_\sun$ (extrapolated from $J$-band and $H$-band data). This value is fully consistent with those cited by \citet{thöne07} and references therein.}
\tablenotetext{e}   {X-ray flash.}
\tablenotetext{f}   {Because of the low Galactic latitude ($b$ = $-4\fdg6$) of this host, we correct for dust-extinction overestimates. Following the recommendation by \citet{dutra03}, we scale the \citet{schlegel98} reddening value multiplying by 0.75 and adopt $E_{\rm MW}(B-V)$ = 0.78\,mag. The UV flux density error (column \UV) contains the additional 25\% uncertainty estimated by \citet{margutti07}.}
\tablenotetext{g}   {Redshift of the 2dFGRS Public Database (Two Degree Field Galaxy Redshift Survey; {\tt http://www2.aao.gov.au/$\sim$TDFgg/}), archival data for galaxy \object[2dFGRS S173Z112]{TGS173Z112}.}

\tablecomments{
                     Because host positions are well determined from previous broadband imaging, upper limits are quoted at the 2$\sigma$ level, while errors are 1$\sigma$. All (UV, optical, NIR, and MIR) flux densities and magnitudes in this table (including those in the table notes) are corrected for foreground Galactic dust extinction. Corrections to the IRAC wavebands follow \citet{lutz99}. For the UV, optical, and NIR passbands we use the DIRBE/IRAS dust maps \citep{schlegel98}. We adopt a Galactic dust extinction curve $A_\lambda/A_V$, parameterised by $R_{V} \equiv A_V/E(B-V)$, with $R_V$ = 3.1 \citep{cardelli89}.
                     Column (\IRAC) Our photometry of {\em Spitzer}'s IRAC, publicly available, archival data (Section \ref{method}). Channels 1, 2, 3, and 4 correspond to wavelengths of 3.6\,$\mu$m, 4.5\,$\mu$m, 5.8\,$\mu$m, and 8.0\,$\mu$m, respectively.
                     Column (\K)    Interpolated flux densities for the rest-frame $K$ band (Section \ref{method}). The data used were obtained from these references.
                     Column (\mass) $M_\star$ derived (Section \ref{M*}) from the rest-frame $K$-band flux densities listed in column (\K), with $M_\star/L_{K_{\mathrm{rest}}}$ = 0.4\,$M_\sun$/$L_\sun$.
                     Column (\UV)   Interpolated flux densities for the rest-frame UV continuum (Section \ref{method}). The data used were obtained from these references.
                     Column (\SFR)  Dust-uncorrected UV SFRs derived (Section \ref{SFRs}) from the rest-frame UV continuum flux densities listed in column (\UV).
\\
              }

\tablerefs{
          $\ddagger$                     This work;
           \citepalias{bloom01}         \citealt{bloom01};
           \citepalias{bloom98}         \citealt{bloom98};
           \citepalias{djorgovski01}    \citealt{djorgovski01};
           \citepalias{bloom99}         \citealt{bloom99};
           \citepalias{tinney98}        \citealt{tinney98};
           \citepalias{djorgovski03}    \citealt{djorgovski03};
           \citepalias{djorgovski98}    \citealt{djorgovski98};
           \citepalias{christensen05}   \citealt{christensen05};
           \citepalias{bloom03}         \citealt{bloom03};
           \citepalias{le floc'h02}     \citealt{le floc'h02};
           \citepalias{piro02}          \citealt{piro02};
           \citepalias{price02a}        \citealt{price02a};
           \citepalias{price02b}        \citealt{price02b};
           \citepalias{price03}         \citealt{price03};
           \citepalias{barth03}         \citealt{barth03};
           \citepalias{jakobsson05b}    \citealt{jakobsson05b};
           \citepalias{vreeswijk06}     \citealt{vreeswijk06};
           \citepalias{maiorano06}      \citealt{maiorano06};
           \citepalias{hjorth03}        \citealt{hjorth03};
           \citepalias{jakobsson04}     \citealt{jakobsson04};
           \citepalias{rau05}           \citealt{rau05};
           \citepalias{prochaska04}     \citealt{prochaska04};
           \citepalias{wiersema04}      \citealt{wiersema04};
           \citepalias{soderberg06}     \citealt{soderberg06};
           \citepalias{pellizza06}      \citealt{pellizza06};
           \citepalias{foley05}         \citealt{foley05};
           \citepalias{pian06}          \citealt{pian06};
           \citepalias{colless01}       \citealt{colless01};
           \citepalias{della valle06a}  \citealt{della valle06a};
           \citepalias{chary02}         \citealt{chary02};
           \citepalias{castro cerón06}  \citealt{castro cerón06};
           \citepalias{le floc'h06}     \citealt{le floc'h06};
           \citepalias{vreeswijk99}     \citealt{vreeswijk99};
           \citepalias{le floc'h03}     \citealt{le floc'h03};
           \citepalias{gorosabel03a}    \citealt{gorosabel03a};
           \citepalias{gorosabel03b}    \citealt{gorosabel03b};
           \citepalias{masetti05}       \citealt{masetti05};
           \citepalias{wainwright07}    \citealt{wainwright07};
           \citepalias{gorosabel05a}    \citealt{gorosabel05a};
           \citepalias{gorosabel05b}    \citealt{gorosabel05b};
           \citepalias{rau04}           \citealt{rau04};
           \citepalias{malesani04}      \citealt{malesani04};
           \citepalias{della valle06b}  \citealt{della valle06b};
           \citepalias{kocevski07}      \citealt{kocevski07};
           \citepalias{thöne08}         \citealt{thöne08};
          (\privatehjorth)                       J. Hjorth (2008, private communication);
           \citepalias{cobb06}          \citealt{cobb06};
           \citepalias{christensen04}   \citealt{christensen04};
           \citepalias{bloom02}         \citealt{bloom02};
           \citepalias{michałowski09}   \citealt{michałowski09};
           \citepalias{margutti07}      \citealt{margutti07};
           \citepalias{sollerman06}     \citealt{sollerman06};
           \citepalias{mangano07}       \citealt{mangano07}.
\\
           }

\end{deluxetable*}

\section{Data}
  \label{data}

Our current sample is composed of 30 long-duration GRB host galaxies, three of them within the X-ray flash category \citep{heise03}. We made the selection by requiring each host to have rest-frame $K$-band data available (for the purposes of this work, we define $K$ $\equiv$ 2.2\,$\mu$m $\pm$ 0.3\,$\mu$m), thus the $M_\star$ estimator is well calibrated \citep{glazebrook04}. An additional requirement for inclusion in the sample was the availability of the redshift.

\begin{figure}
      \epsscale{1.0}
               \plotone{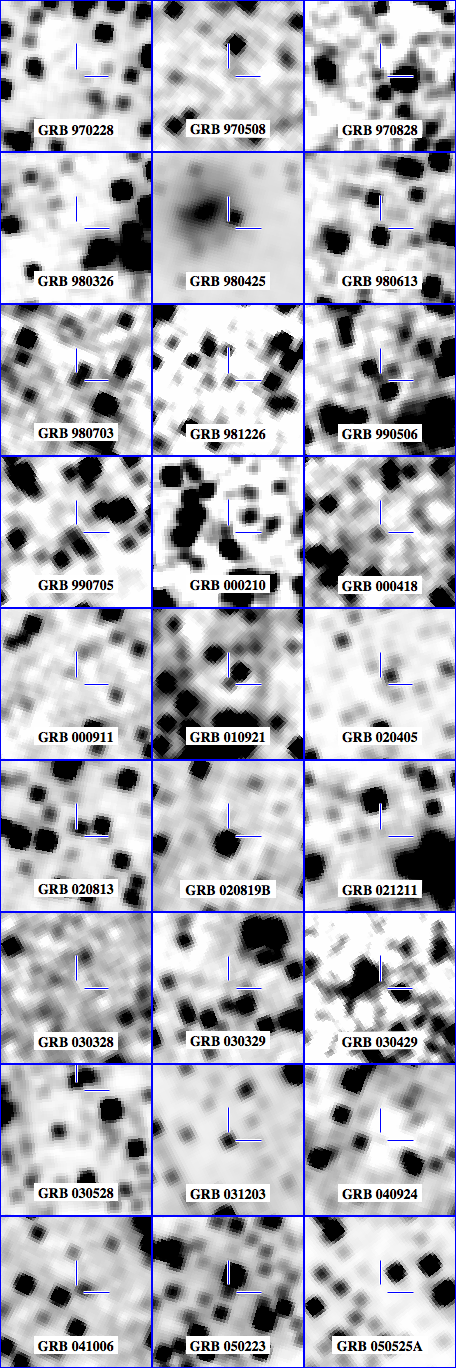}
                       \caption{Postage stamps of the 27 GRB host galaxies included in our sample and for which {\em Spitzer}'s IRAC observations were utilised. Each stamp exhibits a region of 1\arcmin $\times$ 1\arcmin. The ticks mark the exact astrometrically defined position of the GRB. For every case, we employed the best set of coordinates available in the literature. North is upward and east is leftward.
                         \label{mosaic}
                       }
  \end{figure}

The sample of these 30 GRB host galaxies spans a redshift interval 0 $<$ $z$ $<$ 2.7, with a median value $z$ $\simeq$ 0.84. For comparison, the median redshift\footnote{{\tt http://raunvis.hi.is/$\sim$pja/GRBsample.html}} of those GRBs detected prior to the start of operations of the {\em Swift} satellite \citep{gehrels04} is $\langle z \rangle$ $\simeq$ 1.0, and that of those GRBs detected afterward is $\langle z \rangle$ $\simeq$ 2.2; i.e., in this work we are chiefly looking at the lower end of the GRB redshift distribution. Given the redshifts sampled, the rest-frame $K$-band data for 24 of the 30 host galaxies were obtained from the {\em Spitzer} Science Archive, where we examined all publicly available hosts up to (and including) October 2007. The remaining six GRB hosts (\object[GRB980425]{980425}, \object[GRB030329]{030329}, \object[GRB031203]{031203}, \object[GRB060218]{060218}, \object[GRB060505]{060505}, and \object[GRB060614]{060614}) in the sample have very low redshifts ($z$ $\la$ 0.1), so in those cases $K_{\lambda_{\rm obs}}$ $\sim$ $K_{\lambda_{\rm rest}}$ (i.e., $K_{\lambda_{\rm rest}}$ falls within the nominal width of $K_{\lambda_{\rm obs}}$). But for three of these host galaxies {\em Spitzer} data were available, and such data were included in the computation of the rest-frame $K$-band flux density by means of linear interpolation in log space with the corresponding $K_{\lambda_{\rm obs}}$ data (see Section \ref{method}). The sample is presented in Table \ref{flux:m*:sfr}. Figure \ref{mosaic} displays postage stamps for each GRB host galaxy observed by {\em Spitzer}.

Each host (except GRBs \object[GRB060218]{060218}, \object[GRB060505]{060505}, and \object[GRB060614]{060614}) was imaged with IRAC. Detectors are 256 $\times$ 256 square pixel arrays (scale = 1\farcs2\,pixel$^{-1}$ $\times$ 1\farcs2\,pixel$^{-1}$; field of view = 5\farcm21\, $\times$ 5\farcm21\,). The instrumental point-spread functions (PSF; FWHM) are 1\farcs66\,pixel$^{-1}$, 1\farcs72\,pixel$^{-1}$, 1\farcs88\,pixel$^{-1}$, and 1\farcs98\,pixel$^{-1}$ for channels 1, 2, 3, and 4, respectively. The optical and NIR data complementing IRAC in Table \ref{flux:m*:sfr} were obtained from the literature. Two UV data points (GRBs \object[GRB980425]{980425} and \object[GRB060505]{060505}) come from our analysis of {\em GALEX} \citep[{\em Galaxy Evolution Explorer}; ][]{martin03,martin05} data.

\section{Methodology}
  \label{method}

For the MIR photometry, we use official {\em Spitzer} Post Basic Calibrated Data (Post-BCD) products. To ensure the validity of this photometry, we have performed a complete reduction of the corresponding BCD data for a subset of 14 sources (including the six in \citealt{castro cerón06}), carefully following the recommended calibration steps in the IRAC Instrument Handbook. The fluxes resulting from the two methods for these 14 sources were always consistent within errors. The typical discrepancy between the two flux measurements was in the range of 0.1$\sigma$--0.5$\sigma$. We achieved host extraction (see Figure \ref{mosaic}) by combining the world coordinate system (WCS) calibration of the {\em Spitzer} images with the best set of published coordinates for each host galaxy. {\em Spitzer}'s IRAC pointing reconstruction is typically $\la$1\arcsec, and the positions of our GRB hosts were always known a priori to an accuracy of 0\farcs6 or better. The median separation between the host centroid in each IRAC image and the published coordinates is well below 1\arcsec. We checked the astrometric coincidence of the {\em Spitzer} WCS against the best published positions and derived a range of astrometric separations for our objects of 0.6$\sigma$--1.4$\sigma$. \object[GRB980425]{GRB 980425} is the only host galaxy resolved in the IRAC images, and we have obtained its photometry from the literature \citep{le floc'h06}. None of the other GRB host galaxies of our sample are spatially resolved in the IRAC images, and their flux densities can be estimated using small circular aperture photometry. We measure the flux densities over a circled area of radius 2\,pixels. In most cases, this allows us to recover the emission of the host while avoiding contamination from other field sources located nearby. But in a few instances, there was suspicion that the nearby field sources might be contaminating our host galaxy photometry. As a sanity check, we subtracted those field sources and redid the photometry. Field-source subtraction was performed using the detection output image given by the Source Extractor software package \citep[SExtractor;][]{bertin96}, where the detected sources were replaced by background noise. This procedure was repeated several times, randomly varying the seeds to generate the noise in each case. The photometry on the field-source-subtracted images was always consistent with the original aperture photometry. The departure of the flux values (about 0.3--0.4 $\mu$Jy) between the subtracted and non-subtracted photometries was 0.1$\sigma$. This uncertainty was accounted for in the errors listed in Table \ref{flux:m*:sfr}, column (\IRAC), yet it is small in comparison with the intrinsic photometric errors which dominate. Aperture corrections have been applied to account for the extended size of the PSF. We utilised the {\em Spitzer} Science Centre (SSC) recipe for estimating signal-to-noise ratio of a Point Source Measurement for IRAC\footnote{This is no longer available in the SSC Web site, but {\tt http://ssc.spitzer.caltech.edu/warmmission/propkit/som/ irac\_memo.txt} offers a similar recipe adapted for warm IRAC observations.} as a starting point to calculate conservative errors, including both statistical and systematic estimates. Our flux density measurements and upper limits are given in Table \ref{flux:m*:sfr}, column (\IRAC). We find that, of those hosts in our sample observed with channel 1, about 36\% are detected. For channel 2 the rate is about 64\%. This is roughly of the same order as the detection rate by \citet{le floc'h06} with IRAC channel 2 (44\%), though we caution that both samples are incomplete and suffer from selection biases.

For each GRB host, we compute the flux density at the central wavelength of the rest-frame $K$ band by linear interpolation in log space. We interpolate between the IRAC channel and the closest passband, blueward of IRAC, for which data are available in the literature. In a few cases the IRAC waveband corresponds to a rest-frame wavelength shorter than $K$ band, thus we extrapolate linearly. The rest-frame $K$-band flux densities are shown in Table \ref{flux:m*:sfr}, column (\K), along with the appropriate references. In those cases for which only an upper limit to $M_\star$ can be computed, we also estimate a conservative lower limit by linearly extrapolating a flat spectrum ($f_\nu$ $\propto$ $\nu^0$) from the reddest NIR/optical detection available (references in Table \ref{flux:m*:sfr}, column \K). These lower limits are presented as solid bars in Figures \ref{z_m*}, \ref{ssfr_vs_m*}, and \ref{galaxy_samples}.

In the same fashion described above for the rest-frame $K$ band, we compute for each host the rest-frame UV continuum \citep[2\,800\,\AA;][]{kennicutt98} flux density. We either interpolate linearly in log space between the two closest passbands that bracket the rest-frame UV continuum, with data available from the literature, or, when all available data fall redward of 2\,800\,\AA, we extrapolate linearly. These results and the uncertainties resulting from the interpolation or extrapolation are shown in Table \ref{flux:m*:sfr}, column (\UV). The presence of dust curves the spectral shape around 2\,800\,\AA, yet in our case the spectral ranges for interpolating or extrapolating are small \citep{michałowski08}, so they can be well approximated by a linear spectral shape.

All flux densities listed in Table \ref{flux:m*:sfr} are corrected for foreground Galactic dust extinction (see the Notes of Table \ref{flux:m*:sfr} for the details). Conversion of the magnitudes obtained from the literature to flux densities is based on \citet{fukugita95} for the optical passbands and on \citet{tokunaga05} and \citet{cohen03} for the NIR passbands. The error introduced by the assumption of these photometric systems never dominates the photometric uncertainties itself and is safely neglected.

\begin{figure}
      \epsscale{1.15}
               \plotone{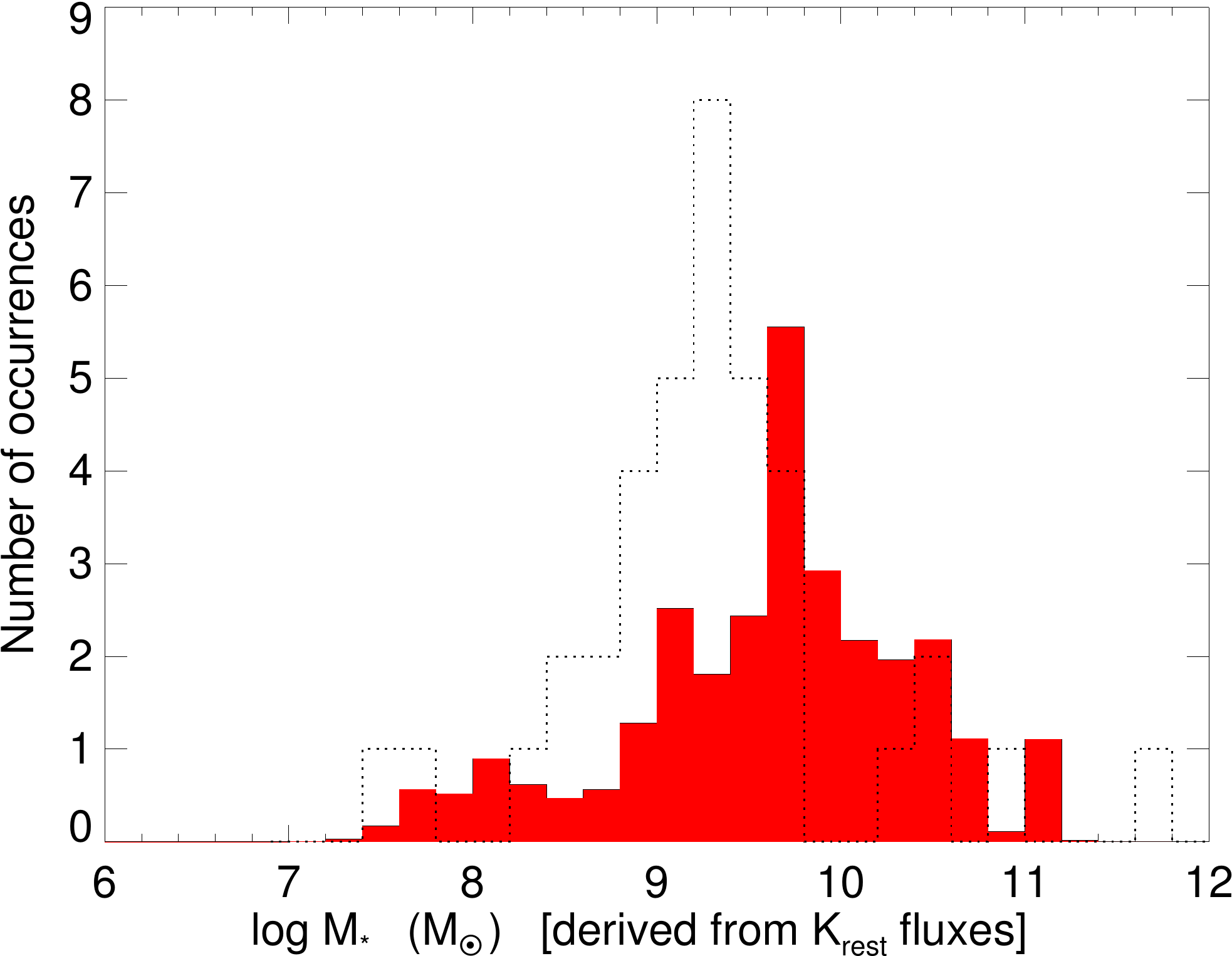}
                       \caption{Distribution of the total stellar mass ($M_\star$) in GRB host galaxies. {\em Filled histogram}: 29 out of the 30 hosts in our sample, spanning a redshift interval 0 $<$ $z$ $<$ 1.5. We note that \protect\object[GRB030429]{GRB 030429} has been excluded from the histogram above and the calculation of the median $M_\star$. This is because its host was never detected at any wavelength and, consequently, no lower limit to $M_\star$ can be estimated. The horizontal axis shows the inferred host $M_\star$, derived from interpolated rest-frame $K$-band flux densities. The median $M_\star$ of the sample is $\langle M_\star \rangle$ = 10$^{9.7}$\,$M_\sun$. For those host galaxies for which we have upper limits, we estimate a conservative lower limit by extrapolating a flat spectrum ($f_\nu$ $\propto$ $\nu^0$) from the reddest NIR/optical detection (references in Table \ref{flux:m*:sfr}, column \K); then we split an area normalised to unit among the bins bracketed by the limits. For each GRB host for which we have detections, we assume a normalised Gaussian distribution of the error bars in linear space. Then, we allocate $M_\star$ in proportion to the area of the Gaussian in each bin. {\em Dotted histogram}: results from \citet[][their Figure 2]{savaglio07}, shown here for comparison. \\
(A colour version of this figure is available in the online journal)
                         \label{m*_histogram}
                       }
  \end{figure}

\section{Total Stellar Masses}
  \label{M*}

We infer $M_\star$ for our sample from rest-frame $K$-band luminosity densities. The light emitted by a galaxy in the $K$ band (e.g., the MIR photometry analysed in this work) is closely related to its $M_\star$ and thus it is a reliable estimator \citep{glazebrook04}. It has little sensitivity to dust since the majority of a galaxy's stellar population has moved away from the birth clouds and because the NIR passbands are virtually unaffected by dust extinction. Such derivation of $M_\star$ is more physically meaningful than the optical/UV luminosity; it effectively integrates over the accumulated $M_\star$ and merger history and can only increase with time, in contrast, for instance, to UV light.

In order to obtain $M_\star$ we apply

\begin{equation}
\label{eq_m*}
M_\star(M_\odot) = 2.67\!\times\!10^{-48}\!\times\!\frac{\scriptstyle 4{\pi}D_L^2f_\nu(\nu_{\mathrm{obs}})}{\scriptstyle 1+z}\!\times\!\frac{\scriptstyle M_\star(M_\sun)}{\scriptstyle L_{K_{\mathrm{rest}}}(L_\sun)},
  \end{equation}

\noindent
where for any given object, $D_L$ is its luminosity distance in cm; $f_\nu(\nu_{\mathrm{obs}})$ is its flux density at the observed wavelength in $\mu$Jy; observations should have been made at wavelengths 1.9\,$\mu$m $<$ $\nu_{\mathrm{obs}}$/(1+$z$) $<$ 2.5\,$\mu$m (e.g., this work); and the factor of 2.67 $\times$ 10$^{-48}$ converts the second term in Equation (\ref{eq_m*}) to units of solar luminosity. The third term in Equation (\ref{eq_m*}), $M_\star/L_{K_{\mathrm{rest}}}$, also in solar units, must be estimated for each object.

$M_\star/L_{K_{\mathrm{rest}}}$ depends to some extent on the composition of the stellar population \citep{portinari04} or, according to \citet{labbé05} who used \citet{bruzual03} with a \citet{salpeter55} IMF, on the rest-frame $U-V$ colour, age, and $M_\star$. GRB host galaxies are blue, young, and faint \citep[e.g.][]{le floc'h03,berger03,christensen04}. In \citet{castro cerón06}, $M_\star/L_{K_{\mathrm{rest}}}$ was assumed to be $\sim$0.1\,$M_\sun$/$L_\sun$ to obtain robust lower limits. For this work, we compute $M_\star/L_{K_{\mathrm{rest}}}$ for GRBs \object[GRB980703]{980703}, \object[GRB000210]{000210}, and \object[GRB000418]{000418} with the rest-frame $K$-band flux densities from Table \ref{flux:m*:sfr}, column (\K), and $M_\star$ values derived from stellar population model SED fitting \citep{michałowski08} and obtain the following results: 0.29\,$M_\sun$/$L_\sun$ for \object[GRB980703]{GRB 980703}, 0.63\,$M_\sun$/$L_\sun$ for \object[GRB000210]{GRB 000210}, and 0.43\,$M_\sun$/$L_\sun$ for \object[GRB000418]{GRB 000418}. These results yield a mean $M_\star/L_{K_{\mathrm{rest}}}$ = 0.45\,$M_\sun$/$L_\sun$, consistent with the average $M_\star/L_{K_{\mathrm{rest}}}$ value in \citet{courty07}, and among the lowest $M_\star/L_{K_{\mathrm{rest}}}$ ratios presented by \citet{portinari04} for a \citet{salpeter55} IMF.

It is sensible to calculate an average $M_\star/L_{K_{\mathrm{rest}}}$ because this ratio is nearly constant, with little dependence on the previous SF history. In fact, $M_\star/L_{K_{\mathrm{rest}}}$ varies only by a factor of 2 between extremely young and extremely old galaxy stellar populations \citep{glazebrook04}. So, to be conservative we estimate 0.4\,$M_\sun$/$L_\sun$ in the calculations of $M_\star$ for our host sample. Table \ref{flux:m*:sfr}, column (\mass) summarises our $M_\star$ estimates. Errors quoted are statistical. We present a histogram of the distribution of $M_\star$ in log space for our GRB host sample in Figure \ref{m*_histogram}. \citet{van der wel06} examined redshift-dependent systematics in determining $M_\star$ from broadband SEDs. They found no significant bias for \citet{bruzual03} models with a \citet{salpeter55} IMF. Nonetheless, some caveats ought to be mentioned here to complement the discussion. We note that the strength of our $M_\star/L_{K_{\mathrm{rest}}}$ determination is limited by the fact that we utilise only three hosts. Stellar population model SED fittings are available for them because these GRB host galaxies have radio and submillimeter detections. This might not be always the case for every GRB host. And some systematic uncertainties were introduced in the computation of the $M_\star$ of those three hosts by employing population synthesis models.

For comparison, we plot in the background of Figure 2 the data from the preliminary analysis of \citet[][their Figure 2]{savaglio07}. The two samples have a 25 object overlap. Our results suggest more massive hosts, about half an order of magnitude higher (median $M_\star$ = 10$^{9.7}$\,$M_\sun$ in ours versus median $M_\star$ = 10$^{9.3}$\,$M_\sun$ in \citealt{savaglio07}; both distributions have a 1$\sigma$ dispersion of 0.8 dex, and in both cases the average has the same value as the median). \citet{savaglio06,savaglio07} fit the optical-NIR SEDs of their host galaxy sample together with a complex set of SF histories. We reproduce the median and average $M_\star$ in \citet{savaglio07} with our data set by applying $M_\star/L_{K_{\mathrm{rest}}}$ = 0.2\,$M_\sun$/$L_\sun$ (the Kolmogorov--Smirnov test indicates to a high probability, $p$ $\sim$ 0.99, likely because of the 25 object overlap, that our data set and that of \citealt{savaglio07} come from a population with the same specific distribution). It thus appears that an adjustment by a factor of $\sim$2 to the $M_\star/L_{K_{\mathrm{rest}}}$ ratio might explain the discrepancies in $M_\star$ between our work and that of \citet{savaglio07}. We note that such an adjustment is within the spread of our calculated values (cf. 0.29\,$M_\sun$/$L_\sun$, 0.43\,$M_\sun$/$L_\sun$, and 0.63\,$M_\sun$/$L_\sun$). The fact that we find larger $M_\star$ may also be indicative of underestimated dust extinction in \citet[][see Section \ref{analysis}]{savaglio07}.

Our $M_\star$ are always lower than those of the normal 0.4 $<$ $z$ $<$ 2 galaxies from the Gemini Deep Deep Survey \citep[GDDS;][]{abraham04,savaglio06}. The GDDS is a deep optical-NIR ($K$ $<$ 20.6) survey complete, for the already mentioned redshift range, down to $M_\star$ = 10$^{10.8}$\,$M_\sun$ for all galaxies and to $M_\star$ = 10$^{10.1}$\,$M_\sun$ for star-forming galaxies. In our host sample, at least 70\% of the galaxies have $M_\star$ $<$ 10$^{10.1}$\,$M_\sun$. This comparison clearly highlights the efficiency of the GRB selection technique, against that of traditional high-redshift surveys, to pick low-$M_\star$ galaxies at high redshifts. This is in agreement with results by \citet{conselice05}. By looking at the light concentration, which is a proxy to total mass, they found that GRB hosts were less concentrated and therefore less massive than field galaxies at similar redshifts. However, caveat has it the GDDS is mass selected and naturally looks for the high end mass of the distribution. From Figure \ref{ssfr_vs_m*}, one can infer that the efficiency of GRBs to pick low-$M_\star$ galaxies at high redshifts might decrease when compared against other surveys.

\begin{figure}
      \epsscale{1.16}
               \plotone{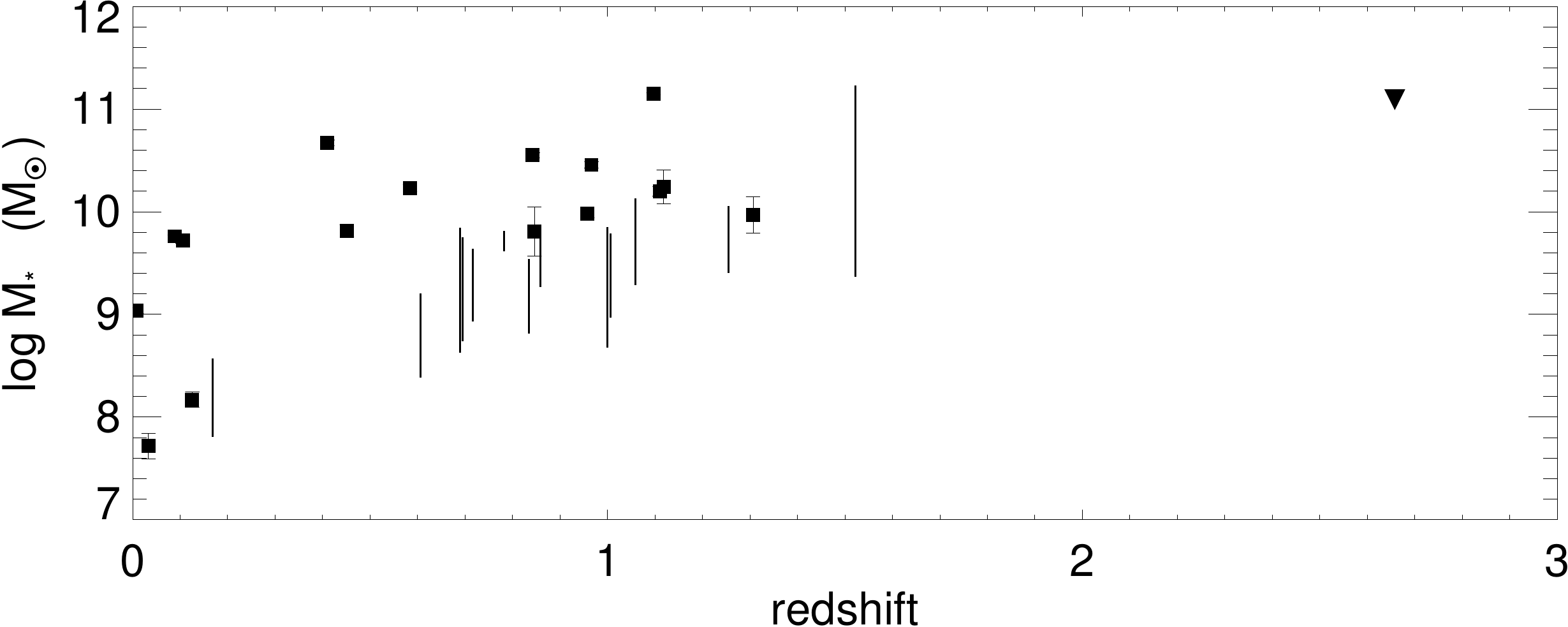}
                       \caption{Total stellar mass ($M_\star$) as a function of redshift for our sample of 30 GRB host galaxies. {\em Squares}: detections. {\em Vertical bars}: distribution of the non-detection measurements of $M_\star$. The lower limits of these bars were calculated by extrapolating a flat spectrum ($f_\nu$ $\propto$ $\nu^0$) from the NIR/optical data (references in Table \ref{flux:m*:sfr}, column \K).
                         \label{z_m*}
                       }
  \end{figure}

We plot $M_\star$ as a function of the redshifts for our sample of 30 GRB host galaxies in Figure \ref{z_m*} and find no intrinsic correlation between the two variables. The scatter of $M_\star$ is rather uniform across most of the redshift distribution. Hosts with very low $M_\star$ are only found at low redshift. For instance, the four GRB hosts (i.e., \object[GRB060218]{060218}, \object[GRB060614]{060614}, \object[GRB030329]{030329}, and \object[GRB980425]{980425}) with the lowest $M_\star$ ($<$ 10$^9$\,$M_\sun$) have some of the lowest redshift values in our sample. Very low-$M_\star$, high-redshift hosts would have been excluded since most of our largely pre-{\em Swift} redshifts were measured in emission, what selects preferentially bright host galaxies. Because the redshift is a requirement for inclusion in our sample, we are effectively biased against faint systems. This situation has now been corrected in the {\em Swift} era when most redshifts are secured via afterglow absorption spectroscopy. The upper limits in the vertical bars of Figure \ref{z_m*} (i.e., the distribution for each non-detection measurement of $M_\star$) mark the sensitivity-limited curve for $M_\star$. Conversely, the absence of high-$M_\star$, low-redshift hosts suggests that such GRB host galaxies are rare.

\section{Star Formation Rates}
  \label{SFRs}

We compute the dust-uncorrected UV SFR for each host by means of their UV continuum luminosity. We convert flux densities into luminosity densities using $L_\nu(\nu_{\rm rest}) = 4{\pi}D_L^2f_\nu(\nu_{\rm observed})/(1+z)$ \citep{hogg02}. Then, we can calculate the dust-uncorrected UV SFRs by applying SFR$(M_\sun\,{\rm yr}^{-1}) = 1.4 \times 10^{-28}L_{\rm UV}$\,(erg\,s$^{-1}$) to the rest-frame $\lambda$ = 2\,800\,\AA\ flux densities \citep{kennicutt98}. The results are summarised in Table \ref{flux:m*:sfr}, column (\SFR). Errors quoted are statistical. In addition, there are systematic errors of order 30\% \citep{kennicutt98}.

\begin{figure}
      \epsscale{1.16}
               \plotone{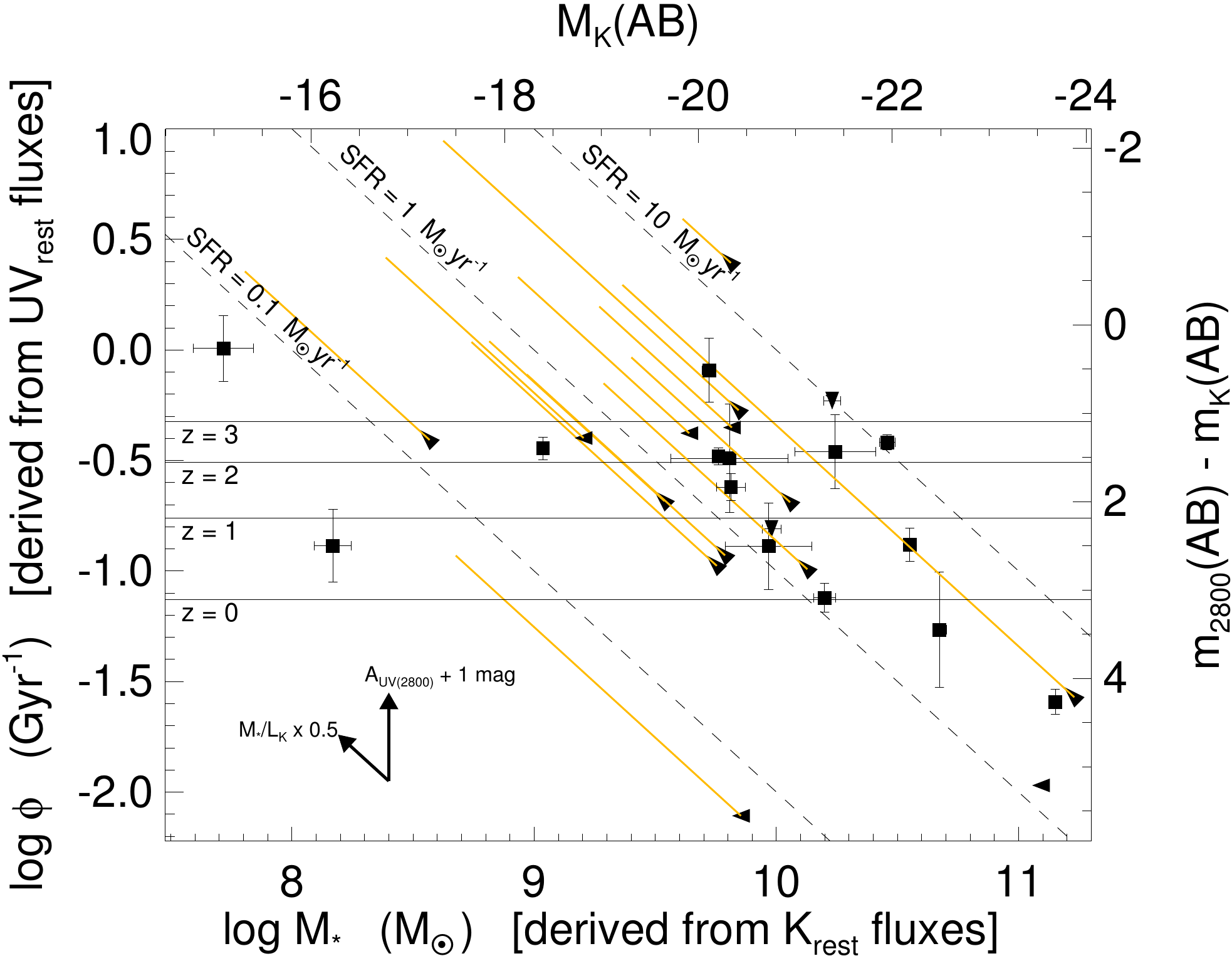}
                       \caption{Specific SFR ($\phi$) as a function of total stellar mass ($M_\star$) for our sample of 30 GRB host galaxies. Squares are detections and triangles mark upper limits for either $M_\star$, SFR, or both. {\em Yellow diagonals}: equivalent to the vertical bars in Figure \ref{z_m*}. Each yellow diagonal could be displaced vertically by the size of the corresponding SFR error bar (Table \ref{flux:m*:sfr}, column \SFR). {\em Dashed diagonals}: constant SFRs of 10, 1, and 0.1 $M_\sun$\,yr$^{-1}$, respectively. {\em Black arrows}: magnitudes of the displacements due to extinction by dust in the UV (e.g., 1 mag; vertical) and changes in our estimation of $M_\star/L_{K_{\mathrm{rest}}}$ $\sim$ 0.4\,$M_\sun$/$L_\sun$ (e.g., a factor of 50\%; diagonal). {\em Right axis}: colour term equivalent to $\phi$. {\em Top axis}: absolute $K$-band AB magnitude, equivalent to $M_\star$. The top/right axes represent our GRB host sample in colour-magnitude space, effectively equivalent to $\phi$ vs. $M_\star$. \\
(A colour version of this figure is available in the online journal)
		                 \label{ssfr_vs_m*}
                       }
  \end{figure}

The specific SFR $\phi$ $\equiv$ SFR/$M_\star$ gives an indication of how intense star-forming a galaxy is. In Figure \ref{ssfr_vs_m*}, we plot $\phi$ versus $M_\star$ for our GRB host sample. The absence of hosts in the lower-left corner is explained as a combination of selection effects and low-number statistics. A host galaxy in this region of the plot has both low $M_\star$ and low SFR, making its detection difficult unless at very low redshifts. As the sampled comoving volume becomes smaller because of the lower redshift required to make a detection, the chance of finding a host decreases accordingly. Given the size of our sample, it is reasonable to expect no detections in this area of the plot. The four GRB host galaxies with the lowest $M_\star$ (GRBs \object[GRB060218]{060218}, \object[GRB060614]{060614}, \object[GRB030329]{030329}, and \object[GRB980425]{980425}) are all at very low redshifts and UV bright. We also note that our sample may be biased against low-SFR hosts, since many redshifts have been measured from emission lines. On the other hand, the non-detection of any GRB host in the upper-right corner of Figure \ref{ssfr_vs_m*} should not be due to a selection effect. Such hosts either do not exist or their afterglows were extincted by dust, thus preventing their localisation. The sample at hand offers some indication as to the former possibility. Our two host galaxies with the highest $M_\star$ (GRBs \object[GRB030328]{030328} and \object[GRB980613]{980613}) would require a dust extinction of $A_V$ $\sim$ 5\,mag to show there. Yet, such dust-extinction levels can be ruled out by the constraints on the SFR from the IR and the radio (see Section \ref{analysis} below).

To exemplify how the estimation of $M_\star/L_{K_{\mathrm{rest}}}$ $\sim$ 0.4\,$M_\sun$/$L_\sun$ affects the location of our hosts in the plot, we suppose a 50\% uncertainty. We repeat the exercise for an UV-continuum dust extinction of 1 mag. The corresponding displacements are plotted in Figures \ref{ssfr_vs_m*} and \ref{galaxy_samples} with black arrows. The magnitude of these displacements is limited enough not to affect our analysis.

\section{Analysis and Discussion}
  \label{analysis}

We find that the GRB host galaxies in our current sample possess a wide range of properties; with 10$^{7}$\,$M_\sun$ $<$ $M_\star$ $<$ 10$^{11}$\,$M_\sun$; and 10$^{-2}$\,$M_\sun$\,yr$^{-1}$ $<$ dust-uncorrected UV SFR $<$ 10\,$M_\sun$\,yr$^{-1}$. Yet, this diversity points toward low $M_\star$, star-forming systems.

Part of our host sample is extinguished by dust. GRB hosts \object[GRB970828]{970828}, \object[GRB980613]{980613}, and \object[GRB990705]{990705} (\citealt[][24\,$\mu$m flux densities]{le floc'h06}; \citealt[][SED fitting]{castro cerón06}), as well as \object[GRB980703]{980703}, \object[GRB000210]{000210}, and \object[GRB000418]{000418} (\citealt[][for detections in the radio and submillimeter wavebands]{berger01,berger03,tanvir04}; \citealt[][for an SED fitting of GRB 980703]{castro cerón06}; \citealt[][for SED modelling]{michałowski08}), have highly obscured SFRs. Additionally, several authors argued for dust extinction in the host of \object[GRB031203]{GRB 031203} \citep[e.g.,][]{prochaska04,margutti07}. Applying the recipe in \citet{castro cerón06} to this host galaxy's MIR photometry \citep[$f_{\rm 3.6\,\mu m}$ = 216\,$\mu$Jy $\pm$ 3\,$\mu$Jy; $f_{\rm 5.8\,\mu m}$ = 390\,$\mu$Jy $\pm$ 16\,$\mu$Jy; $f_{\rm 24\,\mu m}$ = 13\,103\,$\mu$Jy $\pm$ 41\,$\mu$Jy; flux densities corrected for foreground Galactic dust extinction,][]{lutz99,dutra03}, we obtain an SFR$_{L_{\rm 8-1000}}$ = 13\,$M_\sun$\,yr$^{-1}$. That brings the total number of extinguished hosts to at least 7 out of 30 and allows us to crudely estimate that $>$25\% of the sample in this work suffers significant dust extinction ($A_V$ $\ga$ 1\,mag). Neither our sample of host galaxies nor those others cited in this work are bias-free. The searches for the GRBs in such samples have been carried out mostly following the localisation of an optical afterglow, implicitly biasing the sample against dust-extincted systems. Such potential bias strengthens our statement on dust extinction in GRB host galaxies. Parenthetically, we note that GRBs \object[GRB970828]{970828}, \object[GRB980613]{980613}, \object[GRB980703]{980703}, and \object[GRB990705]{990705} make up two-thirds of a redshift-$z$$\sim$1-selected, small subsample \citep{castro cerón06,le floc'h06}. They hint at the possibility that even a higher fraction of hosts are affected by dust extinction, though with the caveat of low-number statistics.

\begin{figure*}
      \epsscale{1.16}
               \plotone{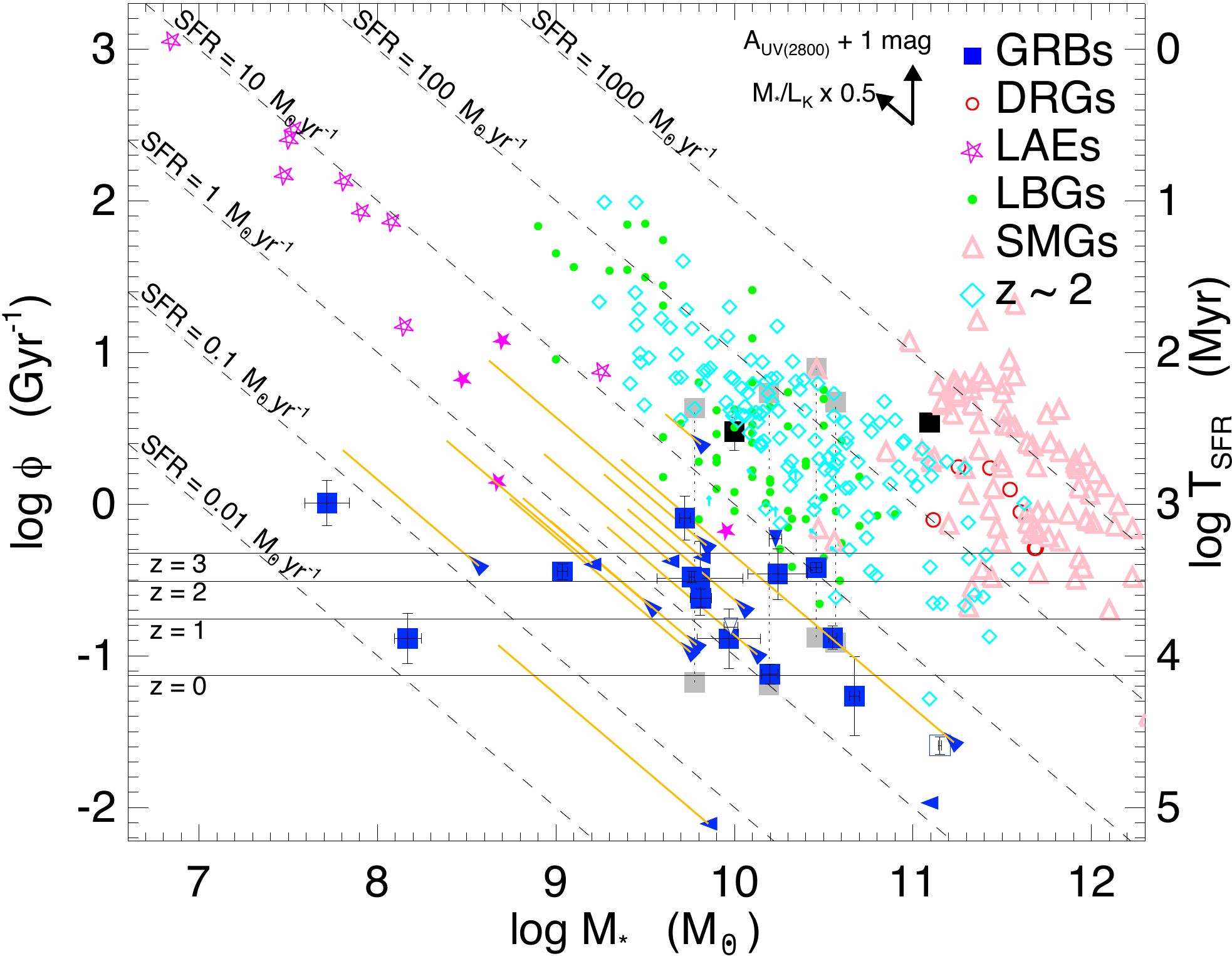}
                       \caption{Specific SFR ($\phi$) as a function of total stellar mass ($M_\star$) for our GRB host galaxy sample and other representative types of galaxies \citep[for a similar plot, see][]{erb06,castro cerón06}. {\em Blue squares and triangles}: host galaxy sample from this work, with dust-uncorrected UV SFRs derived from the rest-frame UV continuum. Triangles mark upper limits for either $M_\star$, SFR, or both. The open blue triangle and square mark the hosts of GRBs \protect\object[GRB970828]{970828} and \protect\object[GRB980613]{980613}, respectively (see Section \ref{analysis}). {\em Black squares}: SFR values constrained with a best-fit SED model \citep{castro cerón06}. {\em Grey squares}: highest/lowest SFR values for those hosts for which a best-fit model could not be established \citep{castro cerón06}. Both the black squares and the grey squares have been shifted here correspondingly to compensate for the difference in $M_\star/L_{K_{\mathrm{rest}}}$ methodology (i.e., from a lower limit value of 0.1\,$M_\sun$/$L_\sun$ in \citealt{castro cerón06} to a best estimated value of 0.4\,$M_\sun$/$L_\sun$ in this work). Yellow diagonals, dashed diagonals, and black arrows are as in Figure \ref{ssfr_vs_m*}. {\em Right axis}: SFR timescale ($T_{\rm SFR}$ = $M_\star$/SFR), the inverse of $\phi$. On this scale, the solid horizontal lines represent the age of the universe for the marked redshift. The distribution of our sample in parameter space suggests that GRBs trace galaxies that are not selected with other techniques. {\em Data points}: GRBs from this work and \citet{castro cerón06}. DRGs from \citet{van dokkum04}. LAEs from \citet{gawiser06}, \citet{nilsson07}, and \citet{lai08}. These are the filled stars; they represent average values of the LAE population as a whole, obtained from stacked photometry. LAEs also from Pirzkal et al. (2007, private communication). These are the empty stars; \citet{pirzkal07} describe these sources but give a single mean value, the result of averaging all of them. Spectroscopically confirmed LBGs from \citet{shapley01} and \citet{barmby04}. SMGs from \citet{borys05} and \citet{chapman05}. From the former we obtained the $M_\star$, from the latter the SFRs, then we have considered $L_{\rm BOL}$ $\simeq$ $L_{\rm far IR}$ and applied the \citet{kennicutt98} calibration. SMGs also from \citet{michałowski10}. $z$ $\sim$ 2 from \citet{erb03} and \citet{reddy06}. \\
(A colour version of this figure is available in the online journal)
                         \label{galaxy_samples}
                       }
  \end{figure*}

\citet{castro cerón06} plotted $\phi$ versus $M_\star$ for six GRB hosts and samples of five other representative types of galaxies: distant red galaxies (DRGs), Ly$\alpha$ emitters (LAEs), Lyman break galaxies (LBGs), submillimeter galaxies (SMGs), and an ensemble of optically selected, $z$ $\sim$ 2 galaxies from the Great Observatories Origins Deep Survey-North (GOODS-N) field. In Figure \ref{galaxy_samples} we plot, with blue symbols, our sample of 30 hosts using a revised $M_\star/L_{K_{\mathrm{rest}}}$ ratio, along with the samples in \citet{castro cerón06}. In our $\phi$ versus $M_\star$ plot (i.e., Figure \ref{galaxy_samples}; we use SFR$_{\rm UV}$), the obscuration of SF by dust pulls the GRB data points down along a vertical line. One way to reconcile the $\phi$ values of host galaxies in \citet{castro cerón06} and this work is to invoke extinction by dust of the order of $A_V$ $\sim$ 1--3\,mag (see below). The conversion from $A_{\rm UV}$ to $A_V$ follows \citet{cardelli89}.

A primary scientific goal in the quantification of galactic evolution is the derivation of the SF histories, as described by the temporal evolution of the SFR($t$). \citet{castro cerón06} noted that their sample had $T_{\rm SFR}$ $<$ $t_{\rm universe}$, allowing for a history of constant SF, with a robust lower limit in $M_\star$ ($M_\star/L_{K_{\mathrm{rest}}}$ $\sim$ 0.1\,$M_\sun$/$L_\sun$). For the sample we present in this work, where we adopt $M_\star/L_{K_{\mathrm{rest}}}$ $\sim$ 0.4\,$M_\sun$/$L_\sun$, clearly a few GRB host galaxies are not allowed to have a history of constant SF (i.e., young stars dominating the stellar populations of old galaxies; see the right ordinate axis in Figure \ref{galaxy_samples}). Either a starburst episode was present in the past or a higher recent SFR is required. The latter possibility is consistent with a fraction of GRB hosts having SF extincted by dust. The hosts of GRBs \object[GRB970828]{970828} and \object[GRB980613]{980613} (open blue symbols in Figure \ref{galaxy_samples}) are good examples because, under the assumption of constant SF, major dust extinction must be invoked to account for the age differences. $\phi_{\rm UV}$ estimates result in $T_{\rm SFR}$ $\sim$ 6\,Gyr for \object[GRB970828]{GRB 970828} and $T_{\rm SFR}$ $\sim$ 32\,Gyr for \object[GRB980613]{GRB 980613}, while $\phi_{\rm IR}$ estimates \citep[see][]{castro cerón06} result in $T_{\rm SFR}$ $\sim$ 300\,Myr for both of them. The discrepancies in $T_{\rm SFR}$ imply a dust extinction of the order of $A_V$ $\sim$ 1.6\,mag for \object[GRB970828]{GRB 970828} and $A_V$ $\sim$ 2.5\,mag for \object[GRB980613]{GRB 980613}. These discrepancies are consistent with the radio-constrained SFR upper limits ($\sim$100\,$M_\sun$\,yr$^{-1}$ for \object[GRB970828]{GRB 970828} and $\sim$500\,$M_\sun$\,yr$^{-1}$ for \object[GRB980613]{GRB 980613}) derived by applying the \citet{yun02} methodology to the deepest radio upper limits reported by \citet{frail03}.

A dilution effect is present in our MIR photometry. Hosts in our current sample are not spatially resolved in the {\em Spitzer} imagery (in the case of \object[GRB980425]{GRB 980425}, we utilise the total flux of the galaxy for consistency with the rest of the sample). To estimate their $M_\star$ we measure the total $K$-band light. $L_K$ traces the accumulation of $M_\star$ \citep{glazebrook04} while, most commonly, the SF is ongoing in only a small part of the host galaxy. So, we do not normalise our sample's dust-uncorrected UV SFRs by the total stellar mass of the star-forming region(s), rather by $M_\star$, which results in lower $\phi$ values. This dilution effect pulls the GRB data points in a $\phi$ versus $M_\star$ plot down along the diagonal (dashed) lines marking constant SFRs.

An apparent envelope can be visualised in Figures \ref{ssfr_vs_m*} and \ref{galaxy_samples}. This is a flat plateau (no objects are present above a certain $\phi$ value, $\sim$2.5\,Gyr$^{-1}$) that starts to curve down beyond a particular $M_\star$ (10$^{10}\,M_\sun$). Extinction by dust, coupled with the dilution effect, could be used to explain this envelope. Correcting for dilution and, chiefly, for dust extinction would yield a new plot where our host sample would align consistently with the results/upper limits of \citet{castro cerón06} and provide support to the claim that GRB host galaxies are small and have some of the highest $\phi$ values.

We conclude by putting forward a simple idea for GRB hosts based on the data analysed here. As a working hypothesis we suggest that, while low $M_\star$ hosts might only contain small amounts of dust (i.e., host galaxies with a low $M_\star$ and a low SFR are rare; see Section \ref{SFRs}), progressing upward in the $M_\star$ distribution of host galaxies will yield significant dust extinction, as well as the already mentioned dilution effect (i.e., the apparent envelope described above for Figures \ref{ssfr_vs_m*} and \ref{galaxy_samples}). Our suggestion is consistent with the theoretical predictions presented in \citet{lapi08}. They further predict that GRB host galaxies trace the faint end of the luminosity function of LBGs and LAEs. Future work (i.e., Herschel observations) on a complete host sample will allow us to test this by quantifying dust extinction and the dilution effect.

The nature of GRBs \object[GRB060505]{060505} and \object[GRB060614]{060614} is strongly debated as no supernova was associated with these long-duration GRBs to deep limits \citep{fynbo06,gehrels06,della valle06a,gal-yam06}. \object[GRB060505]{GRB 060505} falls within the distribution of other long-duration GRB hosts in our sample, whereas \object[GRB060614]{GRB 060614} seems to be an outlier. Though, this may indirectly suggest that the progenitor of \object[GRB060614]{GRB 060614} is different from other typical long-duration GRBs, we note that its SFR is in range with that of the bulk of the sample; and as for $M_\star$, its properties are not very different from some of our other low-redshift host galaxies (e.g., GRBs \object[GRB060218]{060218}, \object[GRB030329]{030329}, and \object[GRB980425]{980425}).

\acknowledgments

We thank Ranga-Ram Chary, \'Ard\'{\i}s El\'{\i}asd\'ottir, Peter Laursen, Bo Milvang-Jensen, and Paul M. Vreeswijk for insightful comments. The Dark Cosmology Centre is funded by the Danish National Research Foundation. J. M. C. C. gratefully acknowledges support from the Instrumentcenter for Dansk Astrofysik and the Niels Bohr Instituttet's International PhD School of Excellence, as well as from the ESA Research Fellowship in Space Science Programme. J. G. was funded in part by Spain's AyA 2.004-01.515 and ESP 2.005-07.714-C03-03 grants. The authors acknowledge the data analysis facilities provided by the Starlink Project which is run by CCLRC on behalf of PPARC. This research has made use of: the NASA's Astrophysics Data System; the GHostS database ({\tt http://www.grbhosts.org/}), which is partly funded by {\em Spitzer}/NASA grant RSA Agreement No. 1287913; the Gamma-Ray Burst Afterglows site ({\tt http://www.mpe.mpg.de/$\sim$jcg/grb.html}), which is maintained by Jochen Greiner; IRAF, distributed by the National Optical Astronomy Observatory, which is operated by the Association of Universities for Research in Astronomy, Inc., under cooperative agreement with the National Science Foundation; the NASA/IPAC Extragalactic Database (NED), which is operated by the Jet Propulsion Laboratory, California Institute of Technology, under contract with the National Aeronautics and Space Administration; and SAOImage DS9, developed by Smithsonian Astrophysical Observatory. {\em GALEX} ({\em Galaxy Evolution Explorer}) is a NASA Small Explorer, launched in April 2003. We gratefully acknowledge NASA's support for construction, operation, and science analysis for the {\em GALEX} mission, developed in cooperation with the Centre National d'Etudes Spatiales of France and the Korean Ministry of Science and Technology.

{\it Facilities:} \facility{{\em Spitzer} (IRAC)} and \facility{{\em GALEX} (NUV)}.

\end{document}